\documentstyle[12pt]{article}

\normalbaselines
\catcode `\@=11
\@addtoreset{equation}{section}

\def\section{\@startsection {section}{1}{\z@}{-3.5ex plus -1ex minus
     -.2ex}{2.3ex plus .2ex}{\normalsize\bf}}
\def\subsection{\@startsection{subsection}{2}{\z@}{-3.25ex plus -1ex
    minus
 -.2ex}{1.5ex plus .2ex}{\normalsize\bf}}

\def\@cite#1#2{${}^{\mbox{\scriptsize#1\if@tempswa , #2\fi}}$}
\def\thebibliography#1{\section*{References\markboth
  {REFERENCES}{REFERENCES}}\list
  {\arabic{enumi}.}{\settowidth\labelwidth{[#1]}\leftmargin\labelwidth
  \advance\leftmargin\labelsep
  \usecounter{enumi}}
  \def\newblock{\hskip .11em plus .33em minus -.07em}
  \sloppy
  \sfcode`\.=1000\relax}
 
\catcode `\@=12

\def\Ref#1{(\ref{#1})}
\def\rz{\ifmmode{I\hskip -3pt R}
    \else{\hbox{$I\hskip -3pt R$}}\fi}
\def\cz{\ifmmode{C\hskip-4.8pt\vrule height5.8pt\hskip6.3pt}
    \else{\hbox{$C\hskip-4.8pt\vrule height5.8pt\hskip6.3pt$}}\fi}
\def\gz{\ifmmode{Z\hskip -4.8pt Z}
    \else{\hbox{$Z\hskip -4.8pt Z$}}\fi}                
\def\grad{\vec{\nabla}\!}                       

\let\ds\displaystyle
\def\dt{\partial_t}                             

\def\F{{\cal F}}

\def\N{{\cal N}}
\def\H{{\cal H}}
\def\J{{\vec{J}}}

\def\Im{\mbox{Im}}
\def\Aff#1{\mbox{\it Aff\/}(#1)}

\begin{document}
\vspace*{2.5cm}
\noindent
{\bf A FAMILY OF NONLINEAR SCHR\"ODINGER EQUATIONS:}\\
{\bf LINEARIZING TRANSFORMATIONS AND RESULTING STRUCTURE}
\vspace{1.3cm}\\
\noindent
\hspace*{1in}
\begin{minipage}{13cm}
H.-D.~Doebner$^{1,2}$, G.~A.~Goldin$^3$, and P.~Nattermann$^{2}$
\vspace{0.3cm}\\
 $^1$ Arnold Sommerfeld Institute for Mathematical Physics\\
 $^2$ Institute for Theoretical Physics,\\
      \makebox[3mm]{ }Technical University of Clausthal \\
      \makebox[3mm]{ }D-38678 Clausthal-Zellerfeld, Germany\\
 $^3$ Departments of Mathematics and Physics, \\
      \makebox[3mm]{ }Rutgers University\\
      \makebox[3mm]{ }New Brunswick, New Jersey 08903, USA \\
\end{minipage}

\vspace*{0.4cm}

\begin{abstract}
We examine a recently-proposed family of nonlinear Schr\"odinger
equations\cite{DoeGol4}
with respect to a group of transformations that linearize
a subfamily of them. We investigate the
structure of the whole family with respect to the
linearizing transformations, and propose a new,
invariant parameterization.
\end{abstract}

\section{\hspace{-4mm}.\hspace{2mm} INTRODUCTION}
Previous work${}^{\mbox{\scriptsize 1--5}}$
on the representation theory of an infinite-dimensional
kinematical algebra on $\rz^{\,3}$,
and the corresponding infinite-dimensional group,
led to a Fokker-Planck type of equation for the quantum-mechanical
probability density and current,
\begin{equation}
\dt\rho = - \grad \cdot \vec{j} + D \Delta \rho\,,
\label{FP:eq}
\end{equation}
and in turn to
a family $\F_D$ of nonlinear Schr\"odinger equations.
$\F_D$ is parameterized by the
classification parameter $D$ of the unitarily inequivalent group
representations (the diffusion coefficient
in Eq. \Ref{FP:eq}),
and five real model parameters $D^{\prime}c_1,\ldots,D^{\prime}c_5$:
\begin{equation}
i\hbar\dt\psi
= \left(-\frac{\hbar^2}{2m}\Delta+V(\vec{x})\right)\psi
+ i\frac{\hbar D}{2} \frac{\Delta\rho}{\rho}\psi
+\hbar D^{\prime}
\left(\,\sum_{j=1}^5 c_jR_j[\psi]\right)\psi\,.
\label{DG:eq}
\end{equation}
Here $D^{\prime}$ also has the dimensions of a diffusion
coefficient (so that the $c_j$ are dimensionless), and
the nonlinear functionals $R_j$ are complex homogeneous of
degree zero, defined by:
\begin{equation}
\begin{array}{c}
\ds  R_1[\psi] := \frac{\grad\cdot\J}{\rho}\,,\qquad
  R_2[\psi] := \frac{\Delta\rho}{\rho}\,,\qquad
  R_3[\psi] := \frac{\J^{\,2}}{\rho^2}\,, \\[1mm]
\ds  R_4[\psi] := \frac{\J\cdot\grad\rho}{\rho^2}\,,\qquad
  R_5[\psi] := \frac{(\grad\rho)^2}{\rho^2}\,,
\end{array}
\label{defineRj}
\end{equation}
where $\rho:=\bar\psi\psi$ and
$\J:=\Im \,(\bar\psi\grad\,\psi) = (m/\hbar)\vec{j}$.

A subfamily of these equations, characterized by
$D^{\prime}c_1 = D = - D^{\prime}c_4$, together with
$c_2 + 2c_5 = 0$ and $c_3 = 0$, satisfies Ehrenfest's theorem
in quantum mechanics\cite{DoeGol4}, and is linearizable via a
nonlinear
transformation\cite{Natter2,AubSab1}. In this
short note we sketch some ideas connected with
the linearizing transformations;
for a detailed derivation and
description with emphasis on the physical
interpretation, we refer to forthcoming
articles\cite{DoGoNa1}.

\section{\hspace{-4mm}.\hspace{2mm} LINEARIZATION}
The members of the Ehrenfest subfamily $\F_D^{\,Ehr}$ can be
transformed into linear Schr{\"o}\-din\-ger
equations by a transformation $\psi\mapsto \psi^\prime=N(\psi)$, if
the
remaining unspecified model parameter
$D^{\prime}c_2$ satisfies
\begin{equation}
\frac{4m}{\hbar}D^{\prime}c_2 \,<\,
1 - \frac{4m^2D^2}{\hbar^2}\,.
\label{inequality}
\end{equation}
Here $N$
depends on two real parameters
$\gamma,\Lambda\in\rz, \Lambda \not= 0$,
and is given by
\begin{equation}
\psi^{\,\prime} := N_{(\Lambda,\gamma)}(\psi) =
\psi^{\,\frac{1}{2}(1+\Lambda+i\gamma)}
{\bar\psi}^{\,\frac{1}{2}(1-\Lambda+i\gamma)} =
|\psi|\, e^{i(\gamma\ln|\psi|+\Lambda\arg\psi)}\,.
\label{gt:eq}
\end{equation}
This maps solutions of the Ehrenfest
subfamily of \Ref{DG:eq} into solutions of the linear Schr\"odinger
equation,
\begin{equation}
i\frac{\hbar}{\Lambda}\partial_t\psi^{\,\prime} =
\left(-\frac{\hbar^2}{2\Lambda^2 m}\Delta
+V(\vec{x})\right)\psi^{\,\prime}\,,
\label{linear}
\end{equation}
for the choices
\begin{equation}
\gamma = -\frac{2mD}{\hbar}\left(1 - \frac{4m}{\hbar}D^{\prime}c_2 -
\frac{4m^2D^2}{\hbar^2}\right)^{-\frac{1}{2}},\quad
\Lambda = \left(1 - \frac{4m}{\hbar}D^{\prime}c_2 -
\frac{4m^2D^2}{\hbar^2}\right)^{-\frac{1}{2}}.
\label{choices:eq}
\end{equation}
However, it should be noted that for non-integer values
of $\Lambda$, the map $N$ is not actually well-defined by
Eq. \Ref{gt:eq} for all $\psi$. The statements in this paper
therefore depend, in some cases, on an appropriate selection
of wave functions.

The transformations $N$ are {\em local\/}, in that they depend
only on the values of $\psi$ and
$\bar\psi$. They also respect the {\em projective\/}
structure of the Hilbert space
$\H=L^2(\rz^{\,3},d^{\,3}\!x)$; i.~e.
for any complex number $c$,
\begin{equation}
N_{(\Lambda,\gamma)}(c\,\psi)= |c|e^{i (\gamma\ln|c|+\Lambda\arg c)}
N_{(\Lambda,\gamma)}(\psi)\,,
\label{rays:eq}
\end{equation}
whence $(c\,\psi)^{\,\prime}$ belongs to the same ray as
$\psi^{\,\prime}$.
Furthermore the transformations leave the symplectic structure $\omega
= \delta\psi\wedge \delta\bar\psi$ on the Hilbert space $\H$ invariant
up to a factor:
\begin{equation}
\begin{array}{rl}
\ds  N_{(\Lambda,\gamma)}^*\omega &\ds = \left(
    \frac{\partial N_{(\Lambda,\gamma)}}{\partial\psi} \delta\psi
    +\frac{\partial N_{(\Lambda,\gamma)}}{\partial\bar\psi}
    \delta\bar\psi \right)\wedge
    \overline{ \left(
    \frac{\partial N_{(\Lambda,\gamma)}}{\partial\psi} \delta\psi
    +\frac{\partial N_{(\Lambda,\gamma)}}{\partial\bar\psi}
    \delta\bar\psi \right)}\\[5mm] \ds
  &\ds = \left(\frac{\partial N_{(\Lambda,\gamma)}}{\partial\psi}
    \frac{\partial \bar N_{(\Lambda,\gamma)}}{\partial\bar\psi} -
    \frac{\partial N_{(\Lambda,\gamma)}}{\partial\bar\psi}
    \frac{\partial \bar N_{(\Lambda,\gamma)}}{\partial\psi}\right)
    \delta\psi\wedge \delta\bar\psi \\[5mm] \ds
  &\ds = \Lambda \omega
\end{array}
\label{symplectic}
\end{equation}
This gives a Hamiltonian formulation for the
Ehrenfest family $\F_D^{\,Ehr}$, as has been
noted elsewhere\cite{DoeGol4,Natter1},
and establishes a connection to the
framework of Weinberg\cite{Weinbe2}.

The set $\N:=\{N_{(\Lambda,\gamma)}\}$
of these nonlinear transformations obeys the group
law of
the affine group $\Aff{1}$ in one dimension,
\begin{equation}
N_{(\Lambda_1,\gamma_1)}\circ N_{(\Lambda_2,\gamma_2)}=
N_{(\Lambda_1\Lambda_2,\,\Lambda_1\gamma_2+\gamma_1)}\,.
\label{affine}
\end{equation}
A further, essential property of $N\in\N$ is,
of course, that it leaves the probability density
invariant:
\begin{equation}
\rho^{\,\prime}(\vec{x},t) = \rho(\vec{x},t)\,,
\label{rt:eq}
\end{equation}
where $\rho^{\,\prime}$ is the density transformed under $N$.
In a certain sense $N$ is
a nonlinear generalization of a linear, $U(1)$-gauge
transformation. We shall call $\N$ the
set of {\em local projective nonlinear gauge transformations\/}.

\section{\hspace{-4mm}.\hspace{2mm} GAUGE INVARIANCE AND
  REPARAMETERIZATION}
The transformation of the current $\J^{\,\prime}$ under $N\in\N$
is given by
\begin{equation}
\J^{\,\prime} :=
\Im\left(\bar\psi^{\,\prime} \grad\,\psi^{\,\prime}\right)
= \rho^{\,\prime}\grad\arg\psi^{\,\prime}
= \Lambda\J + \frac{\gamma}{2} \grad\rho\,.
\label{Jt:eq}
\end{equation}
In order to show the invariance of the family $\F_D$
of equations \Ref{DG:eq},
we rewrite $\F_D$ wholly in terms of densities and currents.
Using the expansion of the Laplacian
$\Delta\psi
= \{iR_1[\psi]+(1/2)R_2[\psi]
- R_3[\psi]-(1/4)R_5[\psi]\}\,\psi$,
we obtain the general form,
\begin{equation}
i\dt\psi = i\sum_{j=1}^2 \nu_j R_j[\psi]\psi
           +\sum_{j=1}^5 \mu_j R_j[\psi]\psi
+ \mu_0 V\psi.
\label{nse:eq}
\end{equation}
{}From \Ref{gt:eq}, \Ref{rt:eq} and \Ref{Jt:eq},
we deduce that
$\psi^{\,\prime} = N_{(\Lambda,\gamma)}(\psi)$
again fulfills \Ref{nse:eq}, but with primed parameters:
\begin{equation}
\begin{array}{c}\ds
\nu_1^\prime = \frac{\nu_1}{\Lambda}\,,\quad
\nu_2^\prime = -\frac{\gamma}{2\Lambda}\nu_1 +\nu_2\,,\\[3mm] \ds
\mu_1^{\,\prime} = -\frac{\gamma}{\Lambda}\nu_1 + \mu_1\,,\quad
\mu_2^{\,\prime} = \frac{\gamma^2}{2\Lambda}\nu_1-\gamma \nu_2
- \frac{\gamma}{2}\mu_1+\Lambda \mu_2\,,\quad \mu_3^{\,\prime} =
\frac{\mu_3}{\Lambda}\\[3mm] \ds
\mu_4^{\,\prime}= -\frac{\gamma}{\Lambda}\mu_3 + \mu_4\,,\quad
\mu_5^{\,\prime} = \frac{\gamma^2}{4\Lambda}\mu_3
- \frac{\gamma}{2}\mu_4
+ \Lambda\mu_5\,,\quad
\mu_0^{\,\prime} = \Lambda \mu_0\,.
\end{array}
\label{pt:eq}
\end{equation}
Thus we have that the
8-parameter
family $\F$ of Eq. \Ref{nse:eq}
is invariant under the action of $\N$; i.e., under the action
of the affine group $\Aff{1}$.
An appropriate description of $\F$ is by means of the {\em orbits\/}
of $\Aff{1}$; since there are 2 group parameters, we look next for 6
functionally independent parameters that are invariant under
$\Aff{1}$. These are the gauge invariants.
After some calculations, we obtain gauge-invariant
parameters:
\begin{equation}
\begin{array}{c}\ds
\iota_1 = \nu_1\mu_2 -\nu_2\mu_1\,,\quad
\iota_2 = \mu_1-2\nu_2\,,\quad
\iota_3 = 1 + \mu_3/\nu_1\,,\quad
\iota_4 = \mu_4-\mu_1\mu_3/\nu_1\,,\\[3mm]\ds
\iota_5 = \nu_1(\mu_2+2\mu_5)-\nu_2(\mu_1+2\mu_4)
    +2\nu_2^2\mu_3/\nu_1\,,\quad
  \iota_0 = \nu_1\mu_0\,.
\label{gi:eq}
\end{array}
\end{equation}
Now if we like we
can choose $\nu_1$ and $\mu_1$ as our group parameters
($\nu_1\neq 0$);
this condition is fulfilled for all modifications of the linear
Schr\"odinger equation, as $\nu_1$ derives from the Laplacian in the
Schr\"odinger equation. Inverting \Ref{gi:eq}, we have
\begin{equation}
\begin{array}{c}\ds
\nu_2 = \frac{1}{2}(\mu_1-\iota_2)\,,\quad
\mu_2 = \frac{1}{2}\nu_1^{-1}(2\iota_1  -\iota_2\mu_1
+\mu_1^2)\,,\\[3mm] \ds
\mu_3 = (\iota_3-1)\nu_1\,,\quad
\mu_4 =  \iota_4 - \mu_1 + \iota_3 \mu_1\, \\[3mm]\ds
\mu_5 = \frac{1}{2}\nu_1^{-1}\left(\iota_5-\iota_1
+ \iota_4(\mu_1-\iota_2) + \frac{1}{2}(\mu_1^2- \iota_2^2)(\iota_3-1)
\right)\,,\quad
\mu_0 = \nu_1^{-1}\iota_0\,.
\end{array}
\label{inverse:eq}
\end{equation}
With this reparameterization the family
of nonlinear Schr\"odinger equations is foliated in leaves
characterized by $\iota_0,\ldots,\iota_5$, of subfamilies
depending on the two group parameters $\nu_1,\mu_1$.
The group of
nonlinear gauge transformations $\N$ acts effectively on each leaf
of the foliation. Because of \Ref{rt:eq}, the
time-evolving probability density for all
points $(\nu_1,\mu_1)$ in a given leaf is the same.

Let us identify the gauge invariants for some special leaves:
\begin{itemize}
\item[a.] The linear Schr\"odinger equation corresponds to the values
$\nu_1 = -\hbar/2m$, $\mu_2 = - \hbar/4m$,
$\mu_3 = \hbar/2m$, $\mu_5 = \hbar/8m$,
$\mu_0 = 1/\hbar$, and $\nu_2=\mu_1=\mu_4=0$.
We then have only two
nonvanishing gauge invariants:
\begin{equation}
\iota_1 = \frac{\hbar^2}{8m^2}\,,\quad
\iota_0 =
- \frac{1}{2m}\,\,\,(\mbox{ for } V\not\equiv 0)\,.
\label{lin:eq}
\end{equation}
Note that $\mu_0$ and $\iota_0$ are indeterminate if
$V \equiv 0$.
So $\hbar$ and $m$ (or their quotient, in the free case) are
gauge-invariant quantities for the family of equations.
\item[b.] The Ehrenfest subfamily $\F_D^{\,Ehr}$ corresponds to the
  values
$\nu_1 = -\hbar/2m$, $\nu_2 = D/2$, $\mu_1 = D$,
$\mu_2 = -\hbar/4m + c_2 D^{\prime}$,
$\mu_3 = \hbar/2m$,
$\mu_4 = -D$,
$\mu_5= \hbar/8m - c_2 D^{\prime}/2$,
and $\mu_0 = 1/\hbar$.
Again there are just two nonzero gauge invariants:
$\iota_0$ as before, and
\begin{equation}
\iota_1 \,=\, \frac{\hbar^2}{8m^2}
\,-\,c_2\frac{\hbar D^{\prime}}{2m}\,-\,\frac{D^2}{2}\,.
\label{ehr:eq}
\end{equation}
Here we rediscover the linearization of the
Ehrenfest family, as when the right-hand side
of \Ref{ehr:eq} is positive,
we can introduce
a new constant $\hbar^\prime$ such that a linear
Schr\"odinger equation with $\hbar^\prime$ replacing $\hbar$ is
contained in the orbit.
\item[c.] For the more general,
Galilei (Schr\"odinger) invariant
subfamily\cite{DoeGol4,Natter3} $\F_D^{\,Gal}$ of $\F_D$,
characterized by the conditions $c_1 + c_4 = c_3 = 0$,
we have the values:
$\nu_1 = -\hbar/2m$,
$\nu_2 = D/2$,
$\mu_1 = c_1D^{\prime}$,
$\mu_2 = -\hbar/4m + c_2 D^{\prime}$,
$\mu_3 = \hbar/2m$,
$\mu_4 = -c_1D^{\prime}$,
$\mu_5 = \hbar/8m + c_5 D^{\prime}$,
and $\mu_0 = 1/\hbar$.
Now we get
four nonvanishing gauge invariants, taking independent values:
$\iota_0$, $\iota_1$, $\iota_2$, and $\iota_5$.
%
%
%
\end{itemize}
\section{\hspace{-4mm}.\hspace{2mm} FINAL REMARK}
We have noted that the transformations
linearizing the Ehrenfest subfamily $\F_D^{Ehr}$
of a family $\F_D$ of nonlinear Schr\"odinger equations
can be viewed as
generalizing the usual $U(1)$-gauge transformations,
and act as a
gauge group $\Aff{1}$ on the parameter space of the family.
We have calculated a parameterization by means of the gauge
invariants,
together with the group parameters of
$\Aff{1}$.

In connection with a more physical interpretation of our foliation
of $\F_D$, we quote a remark by Feynman and Hibbs\cite{FeyHib1}:
\begin{quote}\sl
Indeed all measurements of quantum-mechanical systems could
be made to reduce eventually to position and time measurements.
Because of this possibility a theory formulated in terms of
position measurements is complete enough in principle to
describe all phenomena. {(\rm p. 96)}
\end{quote}
If one adopts this point of view,
quantum theories for which the wave functions
give the same probability
density in space and time are ``in principle'' equivalent.
Hence the leaves
of our foliation consist of
sets of ``in principle'' equivalent quantum-mechanical
evolution equations.

\end{document}